\documentclass[twocolumn,prl,twoside,a4paper]{revtex4}
\usepackage{graphicx}
\usepackage{amssymb}
\usepackage{amsmath,bm}

\begin{document}

\title{Two-timescale stochastic Langevin propagation for  classical and quantum optomechanics}

\author{M. J. Akram, E. B. Aranas, N. P. Bullier, J.E. Lang }
\address{Department of Physics and Astronomy, University College London, Gower Street, London WC1E 6BT, United Kingdom}
\author {T.S.Monteiro}
\email{t.monteiro@ucl.ac.uk}
\affiliation{Department of Physics and Astronomy, University College London, Gower Street, London WC1E 6BT, United Kingdom}

\begin{abstract}
Interesting experimental signatures of quantum cavity optomechanics arise 
because the quantum back-action induces correlations between incident  quantum shot noise and the cavity field. While the quantum linear theory of optomechanics (QLT) has provided vital understanding across many experimental platforms, in certain new set-ups it may be insufficient:  analysis in the time domain may be needed, but QLT obtains only spectra in frequency space; and  nonlinear behavior may be present.   Direct solution of the stochastic equations of motion in time is an alternative, but unfortunately standard methods do not preserve the important optomechanical correlations. 
We introduce two-timescale  stochastic Langevin (T2SL)  propagation as an efficient and straightforward method to obtain time traces with the correct correlations.
We show that T2SL, in contrast to standard stochastic simulations, can efficiently simulate correlation phenomena such as ponderomotive squeezing  and reproduces accurately cavity sideband structures on the scale of the applied quantum noise and even complex features entirely
submerged below the quantum shot noise imprecision floor. We investigate nonlinear regimes and find where comparison is possible, that the method agrees with analytical results obtained with master equations at low temperatures and in perturbative regimes.
\end{abstract}

\maketitle 
\section{Introduction}
\begin{figure*}[t!]
\begin{center}
{\includegraphics[width=6.in]{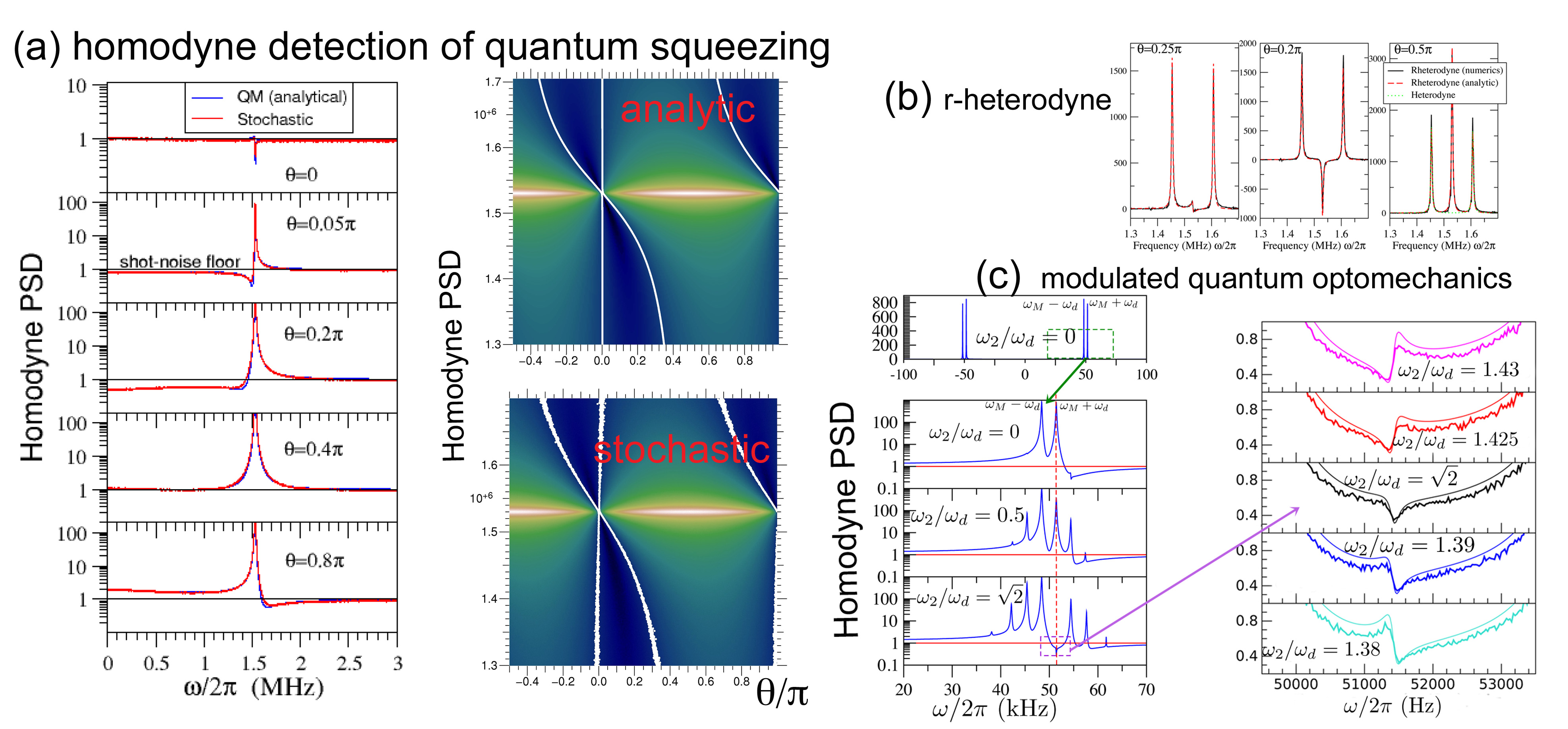}}
\end{center}
\caption{{\bf (a)} Shows T2SL reproduces accurately quantum ponderomotive squeezing effects. Comparison with analytical expressions of quantum linearised 
theory of optomechanics (QLT) is excellent. The presence of correlations $S_{corr}(\omega)$ between optical field quadratures
dependent on the homodyne phase angle result in  PSD values below
the quantum noise floor (black in the colour map on the right). Parameters 
and results are similar to experimentally observed behavior in quantum squeezing regimes seen in
 \cite{Purdy2013}, using both a damping and read-out optical beams and are listed in \cite{Suppinfo}.
Left panels illustrate cuts through the colour map. The shot noise floor $\equiv 1$.
{\bf (b)} Shows T2SL, surprisingly perfectly reproduces sideband structures on the scale of the quantum shot noise kicks. The figure illustrates an example with complex sideband structure completely below the quantum shot noise floor. 
 In levitated experiments with hybrid electro-optical traps both the optomechanical coupling 
$g_1(t) =  2\overline{g} \sin{\omega_\textrm{d}t}$ and $\omega_\textrm{M}(t)=\overline{\omega}_\textrm{M} + 2\omega_2 \cos{(2 \omega_\textrm{d}t)}$ are subject to slow modulations  $\omega_\textrm{d}\ll \omega_\textrm{M}$, resulting in split-sideband peaks, split by $2\omega_\textrm{d}$ at $\pm \omega =\omega_\textrm{M}\pm \omega_\textrm{d}$.
QLT for this case was developed in  \cite{Aranas2016,Aranas2018}:  with increasing $\omega_2/\omega_\textrm{d}$, the $\omega_\textrm{M}+ \omega_\textrm{d}$
peak is suppressed, then re-appears. In squeezing regimes, this second peak is fully submerged below the quantum noise floor: the phase change
around the suppression point is well described by T2SL;
shot noise $\equiv 1$ so  features in the right panels are all below this.}
\label{Fig1}
\end{figure*}

 The field of optomechanics offers a rich arena for investigation of  quantum effects at
mesoscopic or macroscopic scales, on a wide range of experimental platforms including cantilevers, microtoroids, membranes, photonic crystals \cite{AKMreview,FMnoisereview,Bowenbook}.  
Cavity optomechanics allows  manipulation and read-out of the states of small mechanical oscillators via   interaction with a cavity optical mode which is intrinsically nonlinear. The hugely successful quantum linear theory (QLT) of optomechanics has been used to analyse major experimental milestones;
its ease of use and versatility means it is the most widely-used  analytical tool in cavity optomechanics.
 But the importance of nonlinear regimes has also been recognised \cite{Thompson2008,Harris15,Nonlintheory1,Nonlintheory2}  given possibilities  for non-classical state preparation and quantum non-demolition (QND) measurements. 
In certain cases, analysis of nonlinearities via perturbative or quantum master equation approaches may be used; but  this is not always the case as nonlinearities in recent experiments can be large \cite{Verhagen2017}, or  quantum optomechanical effects are seen at higher temperatures  \cite{Sudhir2017,Purdy2017} where the state space is too large. 

  We have shown previously that nonlinearities are important in recent experiments involving  levitated  nanoparticles which exhibit strong nonlinear position coupling to light with couplings 
as $g \propto \cos^2 k{\hat x}$  \cite{Millen2015,Fonseca2016};  other optically trapped set-ups exhibit Gaussian position couplings \cite{Gieseler2012,Rashid2018};  further, the dynamics can evolve in time from strongly nonlinear to linear regimes \cite{Fonseca2016} so temporal analysis is useful. 
New methods  of processing and filtering of the optomechanical experimental time trace (prior to Fourier transforming)  have been demonstrated experimentally \cite{Pontin2018,Setter2018} which further motivate analysis in the time domain. However, QLT does not allow nonlinearity; nor does it yield temporal behaviors  as it directly yields power spectral densities (PSDs)  in frequency space.

Attention is thus turning to explicit solutions of the underlying nonlinear classical or semiclassical stochastic equations of motion
 and these have proved extremely successful for simulation  of
experiments  in cavities \cite{Fonseca2016,Aranas2018} or other optically trapped set-ups \cite{Setter2018,Toros2018},
 for temporal cooling dynamics as well as asymptotic, steady-state regimes.

Solution of the stochastic equations of motion is in principle straightforward and may  be implemented with standard methods and widely available tools \cite{XMDS}.  
What is not  so widely appreciated  is that the normal propagation methods  fail to preserve the vital
important correlations between incoming quantum shot noise $a_{in}$ and the calculated intracavity field $a(t)$, mediated by the backaction.  For fully thermal optomechanical regimes this is not a problem, but otherwise,  including these correlations correctly is essential.  The imprecision quantum noise-backaction correlations underlie  two central signatures of quantum optomechanics: ponderomotive  squeezing of the optical field  and 
Raman sideband asymmetry
  \cite{SideAsymm2012,Khalili2012,Brooks2012,Safavi2013,Purdy2013,Pontin2014,Nunnenkamp2016,Peterson2016}.
The optimal balance between incoming imprecision noise and the back-action component yields the  Standard Quantum Limit (SQL) of force and displacement sensing; squeezing is being investigated as a means to overcome this. 

The failure of the standard solutions to preserve noise correlations is not  because the Langevin equations replace quantum operators by classical complex functions $ \hat {a}(t) \to  a(t)$. These correlations remain relevant even in non-quantum regimes, where their effects are termed noise squashing rather than noise squeezing \cite{Safavi2013a}.

The problem is that while the measured signal obtains the output signal $\hat {a}_{out}(t)$ incorporating correctly all correlations, the calculated signal must rebuild them from an input-output relation such as 
e.g. $\hat {a}_{out}(t)= \hat {a}_{in}- \sqrt{\kappa} \hat {a}(t)$ for a single-sided cavity. Preserving  correlations between the
Markovian shot noise ($\langle  \hat{a}_{in}(t) \hat{a}^\dagger_{in} (t') \rangle = \delta(t-t')$)  and the intracavity dynamics necessitates the complete history of the noise. 

In other words, for a measured signal of length $T$, the {\em experimental} time trace is sampled 
 on a not too small time step $\Delta t$, so the dimension $N=T/\Delta t$  allows for Fourier-transforming, filtering or another further analysis.  
In contrast, for the corresponding {\em theoretical} spectrum of span $T$, the Wiener increments (stochastic noise) are applied on the computational time step $\delta t \ll \Delta t$, so any  trace  with the full noise history is so long
$N_{\delta t}=T/\delta t \gg N$ that even a simple FFT, let alone more complicated temporal analyses e.g. \cite{Pontin2018} become impractical; one may attempt to sample them: this maintains thermal features but would erase the correlation information.

In the present work we propose and test T2SL, a temporal propagation technique that  reproduces these correlations.
The first  step is to separate deterministic and stochastic components of the propagation. The simple premise is that a bath applying much stronger (but far fewer) kicks of variance $\sim \Delta t \gg \delta t$ 
has an equivalent effect. Hence we intersperse deterministic propagation with  noise from a
(still Gaussian) distribution of  higher variance. We can then judiciously pair the $N$ noise contributions to obtain cavity output fields with the correct correlations.

 Comparisons with QLT in linear regimes show the method is accurate and robust enough to easily obtain sideband structures down to the scale of the applied quantum shot noise; in fact it reproduces non-trivial dynamical features completely `submerged' in the quantum shot noise floor and corresponding to near ground-state occupancy. Such features would be considered `beyond the SQL' and would, in experiments, be reported as  significant signatures of quantum optomechanics. It is known that such quantum squeezing can be described by linear theory even semiclassically \cite{Fabre}. However to date, a correlated explicit stochastic demonstration of sidebands `beyond'  the SQL has not been demonstrated. It requires only modest, single desktop computational effort. We then investigate nonlinear regimes and compare with
quantum perturbative methods where possible.

In Sec. 2 we briefly introduce standard optomechanical equations, mainly to define notation, so readers familiar with cavity optomechanics and QLT can move past this section.
In Sec. 3 we introduce T2SL. In Sec. 4 we test the accuracy of the correlated T2SL in a broad range of scenarios to reproduce 
squeezing, including displacement sidebands  weak enough to be below the shot noise floor. Thus the significant and surprising finding here is that the quantum limit of weak shot noise is easily achievable (i.e. that the weaker the sideband, the more challenging the stochastic averaging required to calculate it). We  compare with the results of temporal filtering. Such filtering is of course not possible in QLT. We apply T2SL to the nonlinear regime. To test the method we compare with perturbative nonlinear results obtained in \cite{Nonlintheory1} but  use it to investigate squeezing in the presence of stronger nonlinear coupling, at higher phonon occupancies. Finally in Sec 5 we discuss how the method may complement and augment QLT  in the context  of quantum optomechanics and we conclude.

\section{2. Cavity Optomechanics}

 The simplest optomechanical systems couple a mechanical oscillator to another oscillator corresponding to the optical mode of  a cavity,
and the essential physics is well described by the Hamiltonian:
\begin{eqnarray}
\hat{H}= \Delta \hat {a}^\dagger \hat{a} +
                 \frac{{\hat{\textrm p}}^2}{2m} + \frac{1}{2}m\omega_\textrm{M}^2 {\hat{\textrm x}}^2 +
                 {\hat{H}_{int}} + {\hat{H}_{diss}}
\label{OptoHApp}
\end{eqnarray}
where ${\hat{H}_{int}}$ represents the light matter interaction and ${\hat{H}_{diss}}$ represent dissipative
processes; for the most typical case,
${\hat{H}_{int}}=G_0 \hat {a}^\dagger \hat {a} {\hat{\textrm x}}$ where $G_0$ is the one-photon coupling
strength. To date a linearised analysis, by considering small fluctuations about the
mean $\hat {a} \to \bar{\alpha} + \hat{a}(t)$,  with effective coupling 
  $\hat{H}_{int}=g_1 (\hat {a}^\dagger+ \hat{a}) {\hat{\textrm x}}$ where $g_1\equiv G_0 \bar{\alpha}$ has provided a successful bridge
between theory and experiment in numerous  optomechanics studies.
Nonlinear dynamics arising from position squared coupling 
$H_{int}=g_2(\hat {a}^\dagger+ \hat{a})\hat{x}^2$ can also arise. Unfortunately, nonlinearity (whether optical or position-squared) has proved  difficult to investigate experimentally as it was not previously possible to achieve strong enough  coupling strengths; however new experiments have demonstrated larger or extremely high nonlinear couplings \cite{Painter2015,Verhagen2017}. In addition, at higher temperatures,  experiments with
levitated particles also show $x^2$ dynamics  \cite{Fonseca2016}. 

It is straightforward to generalise dynamical equations to higher numbers of optical field modes and also multiple mechanical oscillators which are all directly or indirectly coupled by the equations of motion, and subjected
to incoming noises (either photon shot noise for the field modes; or incoming phonons for the mechanical modes).
Representing these coupled optical and mechanical modes by a vector $\mathbf X(t) = \begin{pmatrix} \hat a_1 & \hat a^\dagger_1  & \hat a_2 & \hat a^\dagger_2 & \hdots & \hat b_1 & \hat b^\dagger_1 & \hdots \end{pmatrix}^{\textsf T}$, the evolution of this N-dimensional vector, for any instance of linearised optomechanics is given by the form:
\begin{equation}
{\dot {\bf X}}  = {\bf A}(t) {\bf X} + { \bf {\sqrt{\Gamma}}} \zeta(t)
\label{EOM}
\end{equation}
where $ \mathbf {A}(t)$ is an $N \times N$ dimensioned drift matrix, which in general (say modulated optomechanics)
can depend explicitly on time, while ${\bf{\sqrt{\Gamma}}} = \textsf{diag} \begin{pmatrix}
\sqrt \kappa & \sqrt \kappa & \hdots & \sqrt \gamma & \sqrt \gamma & \hdots
\end{pmatrix}$ represents a diagonal matrix of damping coefficients
while $\zeta(t) = \begin{pmatrix}
\hat{a}_{in,1}(t) & \hat{a}_{in,1}^\dagger(t) & \hdots & {\hat b}_{in,1}(t) & \hat{b}_{in,1}^\dagger(t) & \hdots
\end{pmatrix}^{\textsf T}$ represents the (usually Gaussian) noise baths acting on each mode.
The $j$th element of the vector
$(\mathbf {AX})^{(j)}= \frac{1}{i}[\mathbf X^{(j)},{\hat H}]- \frac{1}{2} (\mathbf {\Gamma X})^{(j)}$.
For the simplest case of one-mode and one mechanical oscillator, the individual equation for the read-out (meter) mode (we take $\hat{a}_1 \equiv \hat{a}$ below) becomes simply:
\begin{eqnarray}
{\dot {\hat a}}  =  \left (i \Delta-\frac{\kappa}{2} \right ) {\hat a} + i g \alpha({\hat b}+{\hat b}^\dagger)+ \sqrt{\kappa}{\hat a}_{in}.
\label{meter}
\end{eqnarray}

While the standard linear regime of optomechanics is well established, new types of dynamics arising from 
 modulation of experimental parameters still motivate the development of new approaches to calculation of the spectra
\cite{Fonseca2016,Aranas2016,Aranas2018}.

Generalising the solution of stochastic equations of motion
for the nonlinear case represents a straightforward modification even for multiple interacting modes:
\begin{eqnarray}
{\dot {\bf X}}  = {\bf A}({\bf X},t)  + { \bf \sqrt{\Gamma}} \zeta(t)
\label{EOMnonl}
\end{eqnarray}
and the corresponding equations may be solved by standard packages such as
XMDS \cite{xmds}. Explicit stochastic
propagation of either Eq.\ref{EOM} or Eq.\ref{EOMnonl}
does not show the non-trivial correlations established in the output mode
$ \hat{a}_{out}=\hat{a}_{in}- \sqrt{\kappa}  \hat{a}(t)$  between the
imprecision noises $\hat{a}_{in}$ incident on the cavity and the   
 intracavity field $\hat{a}(t)$ arising from back-action, mediated by the mechanical motion. 

Measurement of the output field by balanced homodyne or heterodyne detection amplifies the signal by beating with a reference oscillator
of phase $\theta$ and obtains a single optical quadrature 
$X_\theta=  \hat{a}(t) e^{-i(\Omega T +\theta)}  +\hat{a}^\dagger(t)e^{i(\Omega t +\theta)}$
where the heterodyne frequency is $\Omega$ and  where $\Omega=0$ for homodyne detection.  \\
The power spectral density (PSD) is
$S_{X_\theta X_\theta}(\omega) = \langle |X_\theta (\omega)|^2\rangle$. 
For QLT, the PSD is obtained by transforming Eq.\ref{EOM} into Fourier space and the 
 averaging of the noise (denoted by the $\langle \  \rangle$)  by substituting noise correlators. The time trace is never calculated.  The measured power spectra density (PSD) however is obtained in frequency space by a Fourier transform (FT) of the time trace
of the measured current. It can also be obtained from an FT of the explicit solutions of Eq.\ref{EOMnonl}.

The PSD can also be expressed in terms of the cavity field components:
\begin{equation}
S_{X_\theta X_\theta}(\omega) = S_{\hat{a} \hat{a}^\dagger}(\omega)+S_{\hat{a}^\dagger \hat{a}}(\omega)
+S_{\hat{a}^\dagger \hat{a}^\dagger}(\omega)e^{2i\theta}+ S_{\hat{a} \hat{a}}(\omega)e^{-2i\theta} \nonumber \\
\label{homo}
\end{equation}
  In other words, the PSD  is the sum of an incoherent part 
and correlations $S_{corr}=S_{\hat{a}^\dagger \hat{a}^\dagger}(\omega)e^{2i\theta}+ S_{\hat{a} \hat{a}}(\omega)e^{-2i\theta}$ (in effect the result of correlations established via the mechanical oscillator between the quadratures of the optical fields). 

Further details are found in \cite{Suppinfo}.

\begin{figure}[ht!]
\begin{center}
{\includegraphics[width=3.3in]{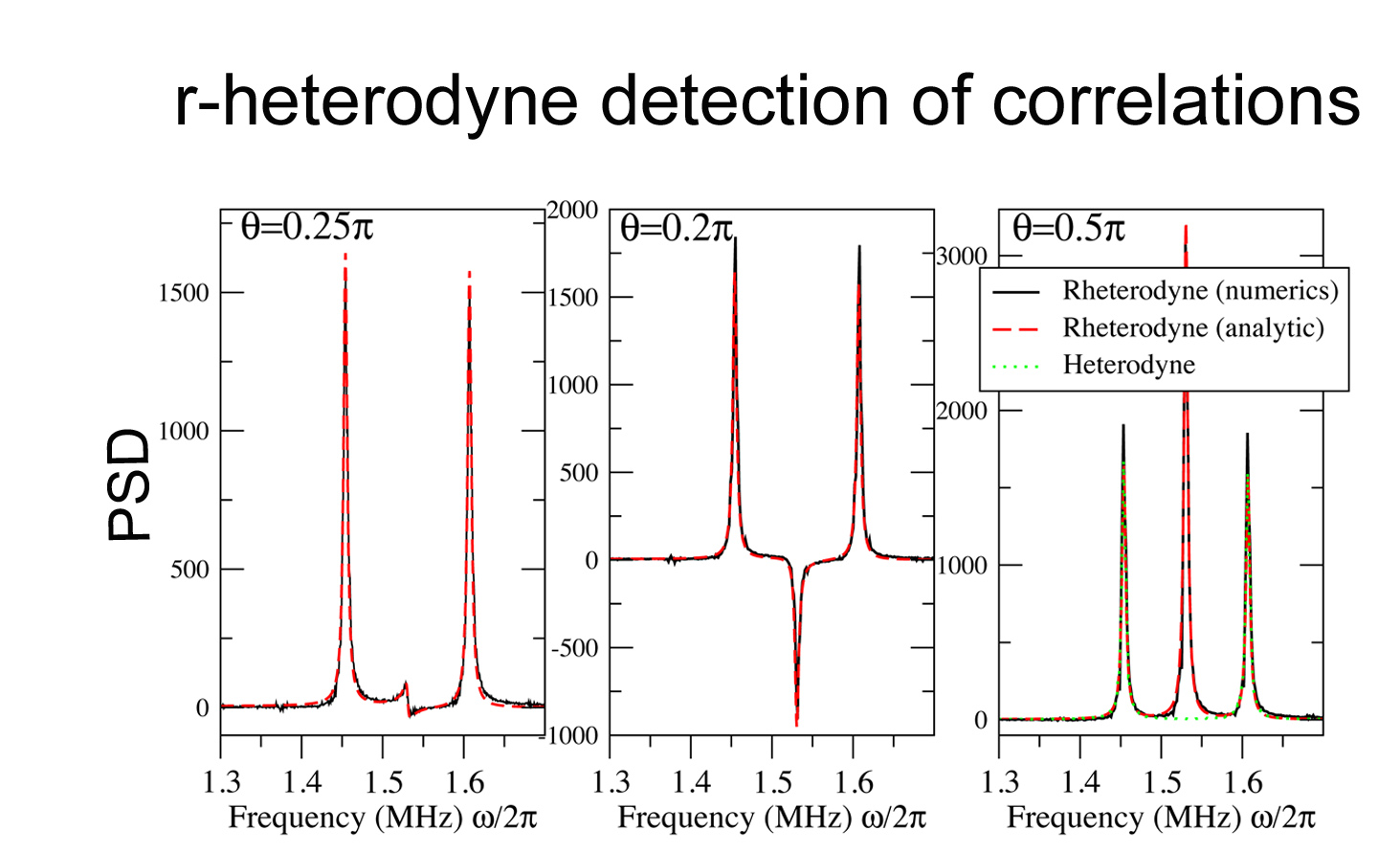}}
\end{center}
\caption{Illustrates the use of filtering in the time domain. T2SL provides a correlated time trace thus allowing temporal processing of the calculated heterodyne or homodyne signal, unlike quantum linearised theory  of optomechanics (QLT) which directly returns spectra in the frequency domain.
Figure illustrates temporal filtering  technique termed r-heterodyne, developed and experimentally tested in \cite{Pontin2018} which extracts homodyne features 
from a heterodyne-detected spectrum using a temporal filter function: a comparison between the
 filtered T2SL time trace and the predicted spectrum 
$S_{\hat{a} \hat{a}^\dagger}(\Omega+\omega)+S_{\hat{a}^\dagger \hat{a}}(\Omega-\omega)
+S_{\hat{a}^\dagger \hat{a}^\dagger}(\omega)e^{2i\theta}+ S_{\hat{a} \hat{a}}(\omega)e^{-2i\theta} $
shows excellent agreement, showing the usefulness of T2SL correlated time traces for allowing the development of new temporal filtering and signal processing methods.}
\label{Fig2}
\end{figure}

\section{3. T2SL}
Numerical propagation of Eq. \ref{EOMnonl} proceeds via an increment:
 \begin{equation}
{\bf X}(t+\delta t)-{\bf X}(t) \equiv d{\bf X}(t) = F({\bf A},\zeta(t))
\label{prop}
\end{equation}
In general, this is constructed via an increment  $F$ which can depend on {\em both} deterministic as well as the stochastic noise terms. Here we start by adopting rather a propagation method which employs independent
increments for the deterministic and stochastic terms such as stochastic Runge-Kutta algorithm for example \cite{Wilkie2004}.  For the optomechanics problems where the stochastic components are weighted by a simple constant matrix, convergence  is robust, so we may write for our increment : $d{\bf X}(t) = R({\bf A})\delta t + d{\bf W}(\delta t)$ where $R({\bf A}) \delta t$ denotes e.g. (deterministic) Runge-Kutta propagation for a time interval $\delta t$, and $d{\bf W}(\delta t)$ represents Gaussian noise baths of zero mean and of variance $\sqrt{\delta t}$ for the optical and mechanical modes.

\begin{figure*}[ht!]
\begin{center}
{\includegraphics[width=6.in]{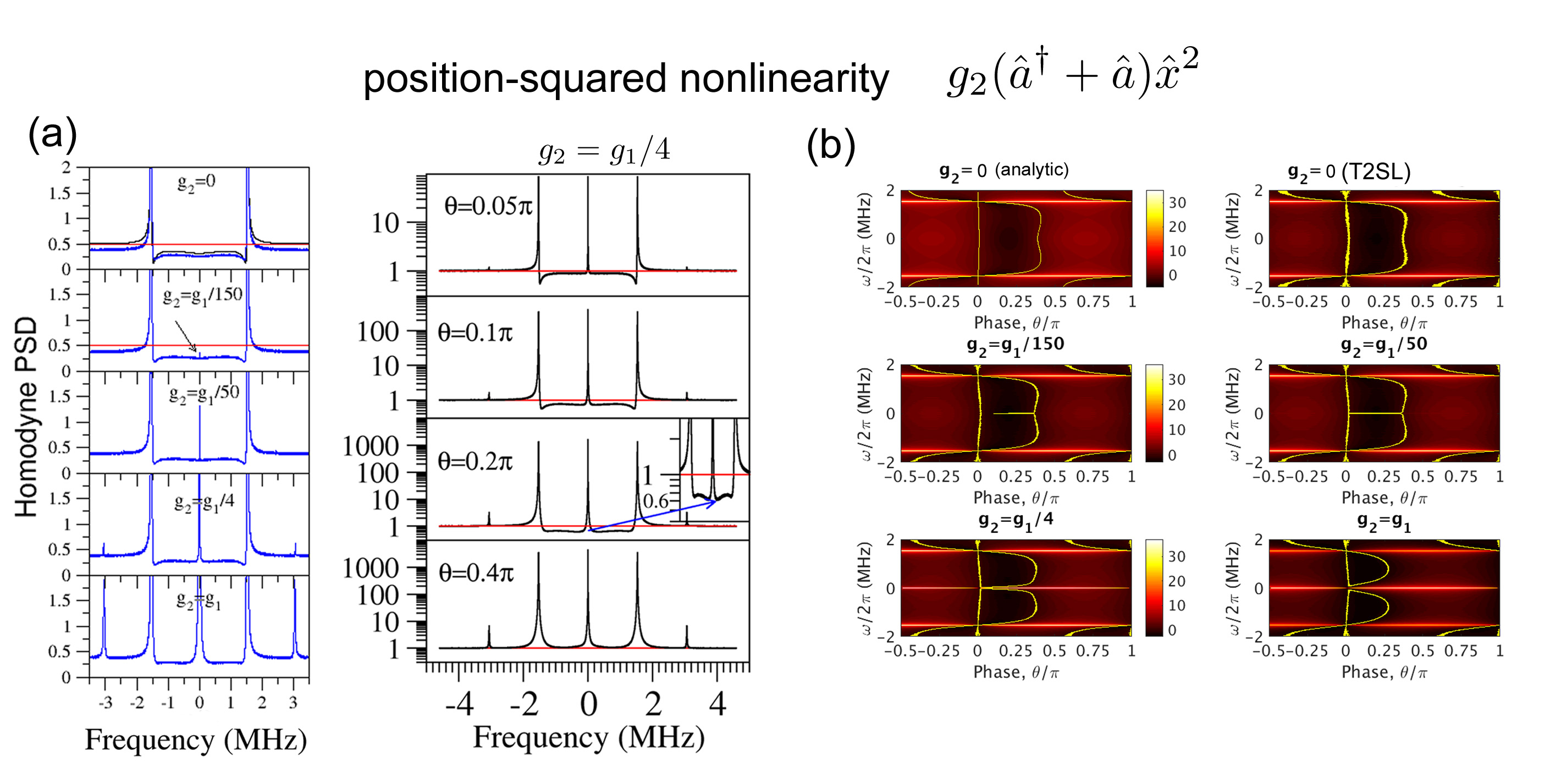}}
\end{center}
\caption{ T2SL for nonlinear quantum optomechanics. T2SL is used in a regime combining linear and position-squared nonlinearity. { \bf (a)}Shows a set of homodyne PSDs (for $\theta=0.05\pi$ 
in the left panels; as a function of $\theta$ in the right panels)
for a system
with optomechanical interaction $H_{int}= g_1(\hat {a}^\dagger+ \hat{a})\hat{x} +  g_2(\hat {a}^\dagger+ \hat{a})\hat{x}^2$ with similar
parameters for Fig\ref{Fig1}(a) apart from the additional nonlinear $g_2$ term \cite{Suppinfo}
The top panel, for $g_2=0$ shows a comparison with QLT. Shows the interaction of ponderomotive 
squeezing with nonlinearity with increasing $g_2$. Note that the central nonlinear peak (left panels) for weaker nonlinearity $g_2 \lesssim 10^{-2} g_1$
can be fully submerged below the quantum shot noise floor. { \bf (b)} shows corresponding 
color maps where black maps the region below quantum shot noise.}
\label{Fig3}
\end{figure*}

However, the problem still remains that to simulate an experimental trace of duration $T$, the number of propagation steps $N_T=T/\delta t \gg 10^7$ is extremely large so preserving the correlations generated by the full history of all these Markovian kicks would still be unfeasible.
Hence for T2SL we introduce a much larger time step $\Delta t$, for the noise kicks.

The first stage is deterministic propagation  for an interval $\Delta t$:
 \begin{equation}
{\bf X}(t+ \Delta t^-)= \int_t^{t+\Delta t^-} {\bf A}({\bf X},t) dt
\label{propdet}
\end{equation}
 where the integral indicates  propagation of the coupled equation of motion by some convenient numerical algorithm (taking many
small time computational time steps $\delta$)  and where $t+ \Delta t^-$ denotes the instant  just before the noise ``kick''.

This is followed by application of the Gaussian noise increments:
 \begin{equation}
{\bf X}(t+ \Delta t)= {\bf X}(t+ \Delta t^-)+ d{\bf W}(\Delta t)
\label{propdet}
\end{equation}
 The correlators of the noise bath (see \cite{Suppinfo}) are unchanged.
this is then repeated, to generate a time series $\{ {\bf X}(t_j= j\Delta t)\}$ for $j=1,2,...N$ from a series of random kicks $d{\bf W}^{(j)}(\Delta t)$,
where $N=T/\Delta t \ll T/\delta t$ for a signal of length $T$.

In the  optomechanical systems of interest we have, for the $j-$th noise kicks at time $t_j$, a vector 
 \begin{eqnarray} 
d{\bf W}^{(j)}=\begin{pmatrix} dW_1^{(j)} & dW_2^{(j)}& \hdots & dW_k^{(j)} & \hdots \end{pmatrix}^{\textsf T}\nonumber\\
= \begin{pmatrix} \sqrt{\kappa(n_p+1/2)}\zeta^{(j)}_p & \hdots & \sqrt{\gamma(n_m+1/2)}\zeta^{(j)}_m) & \hdots \end{pmatrix}^{\textsf T}\nonumber\\
\end{eqnarray}
 of dimension $l=1,2,...2n_{mod}$ where $n_{mod}$ is the number of independent modes (photon and mechanical)
 where the corresponding $\zeta^{(j)}_{p,m}$ (photon and mechanical) are $2n_{mod}$  Gaussian random numbers drawn
from distributions of variance $\sqrt{\Delta t}$.  We note that ${\bf X}(t) = \begin{pmatrix} & \hdots & a(t) & a^\dagger(t) & \hdots & b(t) & b^\dagger(t)) \end{pmatrix}^{\textsf T}$ includes all modes. \\

We focus on the probe beam/read-out mode $a(t)$,  {\em which we take to be the $l-$th element} of ${\bf X}(t)$. 
The corresponding component in Eq.\ref{propdet} 
which propagates the intracavity field of this mode is  $a(t =j\Delta t))= a(t=j\Delta t^-)  + dW_l^{(j)}$.
The above is  standard stochastic numerics, other than the fact that the noise time step (and its variance) is much larger $ \Delta t \gg \delta t$. 
Numerical comparisons show that the intracavity field is insensitive to this modification and excellent agreement with QLT is still obtained
in linear regimes.

The key step in T2SL is the propagation of the output field. We consider the input-output form for a single-sided cavity,
  $ \hat{a}_{out}=\hat{a}_{in}- \sqrt{\kappa}  \hat{a}(t)$ but our method is easily adapted to other cases.
 We propose the following stochastic increments for the correlated output field:
 \begin{eqnarray}
 a_{out}(t_j=j\Delta t)= \mathcal{C} dW_l^{(j)} -\sqrt{\kappa}[a(j\Delta t^-) + dW_l^{(j)}]
\label{propout}
\end{eqnarray}
where $\mathcal{C}$ is a constant. The second term (in square brackets) is  simply the update to the intracavity field while the first term is 
the imprecision noise. We choose the normalisation $\mathcal{C}$ such that the time
series $\{\mathcal{C} dW_l^{(j)} (t_j)\}$ gives a flat noise spectrum of height $n_p+1/2$ (and specifically height $=1/2$ for quantum shot noise for which $n_p=0$). Here we took $\mathcal{C}=1/(\sqrt{\kappa}\Delta t)$ but this can be adjusted for alternative implementations e.g. dependent on  the FFT normalisations. The main point is to set $\mathcal{C}$  such that if  $a_{out}(t_j) = \mathcal{C} dW_l^{(j)}$, the required white noise floor level is obtained. And to ensure that the same noise kick $dW_l^{(j)}$ is used in both terms in Eq.\ref{propout}.
Clearly the term $a(t=j\Delta t^-)$ contains the past history of kicks $1,2,...j-1$ which drive the dynamics and back action. It carries all the narrowband features that interfere with the imprecision floor.

 But we show below this pairing of kicks in Eq.\ref{propout} preserves the important
 correlations so obtains spectral features of both classical and even certain important regimes of {\em quantum}  optomechanics.

\section{4. Applications of T2SL}
We now test and consider applications of T2SL to a variety of regimes including comparisons with 
spectra in quantum regimes.

\subsection{(a) Linear: quantum optical squeezing}

 The resultant explicit stochastic numerics are in excellent agreement with results obtained from analytical quantum noise spectra in linearised regimes.
 Fig.\ref{Fig1} (a) compares the stochastic model and QLT analytical spectra corresponding to recent experiments \cite{Purdy2013} 
on ponderomotive squeezing.  The comparison shows these regimes of phonon occupancies of 10's of quanta 
are easily calculated stochastically with modest averaging of a few 100's of trajectories, taking minutes of CPU 
on a desktop. 

\subsection{(b) Linear: Modulated quantum optomechanics}

Fig. \ref{Fig1}(b) presents  a more much more  challenging test of T2SL, a regime modelling a levitated nanoparticle in a hybrid optical cavity Paul trap in the quantum backaction limit. 
The trapping potential introduces a temporal modulation which introduces non-stationary 
components in the PSD calculations and makes even the QLT much more complicated. The QLT theory was developed in \cite{Aranas2016,Aranas2018} but there was only compared with stochastics in thermal regimes. Further details are also also in the Supplementary Information.

Here we are able to compare the modulated QLT with T2SL calculations. In back-action dominated regimes, squeezing can lead to complicated sideband behavior completely immersed below the quantum imprecision floor. These very small structures, on the order of a single quantum and comparable to the noise are more challenging to obtain. Nevertheless with about 1000 trajectory averaging, a modest calculation yields near exact agreement.

\subsection{ (c) Temporal filtering of correlated signals}
Fig.\ref{Fig2} exemplifies the advantages of enabling direct manipulation of temporal traces prior to the FFTs  which yield PSDs or other types of spectra.  
QLT does not allow this as experimental comparison is only possible in frequency space yet such methods are proving useful in experiments \cite{Pontin2018,Rashid2018}. 
In particular \cite{Pontin2018} demonstrated a temporal technique to restore homodyne-like features arising from correlations which are eliminated in heterodyne detection. In \cite{Pontin2018} a time-filter was applied to the experimental time-trace; then an FFT enabled comparison with the appropriate QLT 
generated components.

Here we show that we can apply the filter directly to the calculated T2SL trace. This means that filtering protocols can be perfected and tested easily without the need for experiments in squeezing regimes which are laborious and difficult.
It would be interesting to apply T2SL to other proposals such as synodyne detection \cite{Buchman} , which involves two heterodyne reference signals, possibly in combination with filtering. 
It may also be applied to recover non-stationary components in such spectra or modulated optomechanics.

\subsection{(d) Nonlinear: quantum optical squeezing}
It is then straightforward to extend the method to systems with nonlinearities, both optical and position squared; in fact any other type of nonlinearity can thus be simulated and the method may also be applied to other general situations including dissipatively coupled systems. In Fig.\ref{Fig3} PSDs showing both squeezing as well as the effects on nonlinearity are shown. In Fig.\ref{Fig3}(a) the ${\hat x}^2$ nonlinearity appears in the form of mechanical sidebands
at both $\omega=0$ and $\omega=2\omega_M$. The central $\omega=0$ detected peak is far stronger; this is a straightforward effect of the cavity filtering since the calculations corresponded to sideband-resolved regimes with 
$\omega_M \gtrsim \kappa/2$ so the $2\omega_M$ is strongly suppressed; but it indicates the advantages of detecting
weak nonlinearities using the (not so well studied) central DC peak, using careful balanced detection.
The clearest advantage is that the effects of squeezing are strong near $\omega=0$ and for a system with  
$g_1 \sim g_2$ the SQL can be overcome at the mechanical frequency itself, unlike the case for pure linear coupling.
All the simulations in Fig.\ref{Fig3} average over  500-1000 stochastic realisations.

\begin{figure}[t!]
\begin{center}
{\includegraphics[width=3.3in]{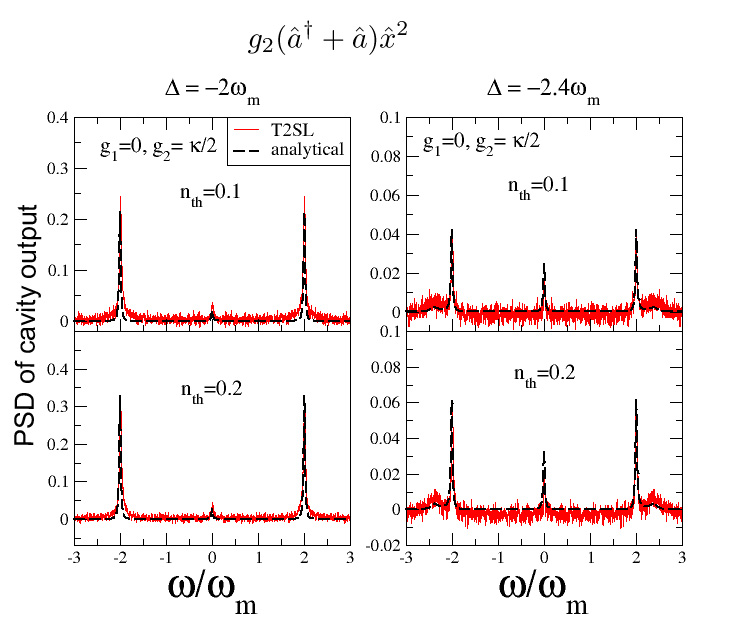}}
\end{center}
\caption{ T2SL for nonlinear quantum optomechanics. T2SL may be employed in regimes of extremely low thermal occupancies $n_{th} \lesssim 1$  of the mechanical oscillator. Here we have pure position-squared nonlinearity (linear coupling $g_1 = 0$) and optomechanical  interaction $H_{int} =g_2(\hat{a} + \hat{a}^\dagger) \hat{x}^2$ and bath temperatures $T_B = 0$ so it is possible to compare with analytical perturbative results \cite{Nonlintheory1} (black line).  The figure shows the PSD of the cavity
output and compares with Eq.7 of  \cite{Nonlintheory1} for two values of detuning. }
\label{Fig4}
\end{figure}

\subsection{(e) Nonlinear: Comparison with quantum perturbative methods}
In Fig.\ref{Fig4} we compare T2SL for a pure nonlinear (position squared)  coupling near zero temperature with analytical expressions obtained perturbatively in \cite{Nonlintheory1}. This was the most challenging case numerically as the coupling is extremely weak and involves an extreme quantum regime of low phonon occupancy. 
It required 10000's of realisations. Although not the most appropriate regime for T2SL it indicates that 
position squared nonlinearities are also adequately captured by T2SL.

\section{5.Discussion: T2SL for quantum optomechanics}

We have shown that T2SL accurately reproduces quantum ponderomtive squeezing regimes.
However, we note that it does not obtain Raman sideband asymmetries another key experimental signature of the quantum optomechanical regime which can arise from the quantum back-action \cite{Weinstein2014,Borke2016}.

It is useful here to divide  optomechanical quantum signatures which arise from correlations between the quantum shot noise and the intracavity field  into two classes: (i) those that are sensitive 
to the commutation relation $\langle [ \hat {a}_{in}(t),\hat {a}^\dagger_{in}(t')]\rangle=\delta(t - t')$ (this includes Raman sideband asymmetry) and (ii) those that are insensitive and yield the same calculated spectra if one assumes 
$\langle  \hat {a}_{in}(t)\hat {a}^\dagger_{in}(t')\rangle= \langle  \hat {a}^\dagger_{in}(t)\hat {a}_{in}(t')\rangle=1/2 \delta(t - t')$ (these include ponderomotive
squeezing and quantum optical correlations such as are obtained by homodyne detection or cross-correlation spectra).
One might argue that only the former are `real' quantum optomechanical signatures while the latter are classical phenomena.
However,  the insensitivity is rather a result of the symmetrisation of the detected spectra; hence ponderomotive
squeezing is an important signature of quantum optomechanics. 

Regardless of such classification, optical quantum squeezing is  at the heart of  current schemes to achieve and overcome the SQL in optomechanical displacement sensing. Although there is currently much interest
in the relation between asymmetry in the mechanical sideband spectrum vis-a-vis the cavity output spectrum \cite{Weinstein2014,Borke2016} for different detection scenarios, here there is no such ambiguity: the mechanical spectrum is always symmetric for these Langevin simulations of quantum squeezing; nevertheless, the total sideband area remains correct so the calculations are still useful for sensing of displacement and forces as well as thermometry. Hence T2SL offers  a new and quite straightforward approach to investigate 
these important regimes for both quantum noise squeezing and classical noise squashing.

{\em Conclusion} We show that by means of a purely classical calculation using T2SL, one can accurately reproduce optomechanical spectra below the Standard Quantum Limit (SQL) in narrowband measurements where the homodyne measurement falls below the shot noise imprecision floor. We have investigated regimes with simultaneous linear and nonlinear couplings. Even in linear regimes, T2SL complements QLT since the method generates traces in the time domain.
It can also be employed in complex linear regimes (multi-mode optomechanics, modulated optomechanics) as an independent check of QLT.
The advantage is that T2SL can then be employed to explore regimes beyond the scope of the usual  linear analysis and temporal filtering.

 \end{document}